\documentclass[a4paper,fleqn,usenatbib]{mnras}

\usepackage{graphicx}	
\usepackage{amsmath}	
\usepackage{amssymb}	
\usepackage{natbib}
\usepackage[flushleft]{threeparttable}
\usepackage{pdflscape}	
\usepackage{soul}

\title[Dust produced by stellar sources]{Where does galactic dust come from?}

\author[M. Ginolfi et al.]{M. Ginolfi$^{1}$\thanks{E-mail: michele.ginolfi@oa-roma.inaf.it},
L. Graziani$^{1}$,
R. Schneider$^{1,2}$,
S. Marassi$^{1}$,
R. Valiante$^{1}$,\newauthor 
F. Dell'Agli$^{3,4}$,
P. Ventura$^{1}$, 
L. K. Hunt$^{5}$
\\
$^{1}$INAF - Osservatorio Astronomico di Roma, via Frascati 33, 00078 Monte Porzio Catone, Roma, Italy\\
$^{2}$Dipartimento di Fisica,  \textquoteleft Sapienza\textquoteright\,  Universit\`{a} di Roma, Piazzale Aldo Moro 5, 00185, Roma, Italy\\
$^{3}$Instituto de Astrofisica de Canarias, Via Lactea, E-38205 La Laguna, Tenerife, Spain\\
$^{4}$Departamento de Astrofisica, Universidad de La Laguna (ULL), E-38206 La Laguna, Spain\\
$^{5}$INAF/Osservatorio Astrofisico di Arcetri, Largo E. Femi 5, 50125 Firenze, Italy}

\date{Accepted XXX. Received YYY; in original form ZZZ}

\pubyear{2017}

\begin{document}
\label{firstpage}
\pagerange{\pageref{firstpage}--\pageref{lastpage}}
\maketitle

\begin{abstract}
Here we investigate the origin of the dust mass ($\rm M_{dust}$) observed in the Milky Way (MW) and of dust scaling relations found in a sample of local galaxies from the DGS and KINGFISH surveys.
To this aim, we model dust production from Asymptotic Giant Branch (AGB) stars and supernovae (SNe) in simulated galaxies forming along the assembly of a Milky Way-like halo in a 
well resolved cosmic volume of 4 cMpc using the \texttt{GAMESH} pipeline. 
We explore the impact of different sets of metallicity and mass-dependent AGB and SN dust yields on the predicted $\rm M_{dust}$.
Our results show that models accounting for grain destruction by the SN reverse shock predict a total dust mass in the MW that is a factor of $\sim4$ lower than observed, and can not reproduce
the observed galaxy-scale relations between dust and stellar masses, and dust-to-gas ratios and metallicity, with a smaller discrepancy in galaxies with low metallicity (12 + log(O/H) $<$ 7.5) and
low stellar masses ($M_{\rm star} < 10^{7} M_\odot$). 
In agreement with previous studies, we suggest that competing processes in the interstellar medium must be at play to explain the observed trends. Our result reinforces this conclusion by showing that 
it holds independently of the adopted AGB and SN dust yields. 

\end{abstract}



\section{Introduction}\label{sec:introduction}

Interstellar dust plays a major role in driving galaxy evolution along cosmic time due to its tight connection with both star formation (SF) and the thermal and chemical evolution of the interstellar medium (ISM).
Dust grains can shield molecules from the photo-dissociating radiation and trigger the formation of molecular hydrogen (H$_2$), acting as an efficient catalyst of atomic hydrogen (HI) reactions. 
In addition, dust shapes the observed galaxy colours by absorbing and scattering stellar light at ultraviolet (UV) and visible wavelengths, and re-emitting infrared (IR) radiation. 

Interstellar dust grains are produced in the circum-stellar envelopes of asymptotic giant branch (AGB) stars (\citealp{Ferrarotti&Gail2006}; \citealp{Zhukovska2008}; \citealp{ventura_etal_2012a, ventura_etal_2012b};   \citealp{dicriscienzo_etal_2013}; \citealp{Nanni2013}; \citealp{ventura_etal_2014}; \citealp{DellAgli2015, DellAgli2017}) and in the ejecta of core-collapse supernovae (SNe, \citealp{Todini&Ferrara2001}; \citealp{Schneider2004}; \citealp{Nozawa2007}; \citealp{Bianchi&Schneider2007}; \citealp{Cherchneff2009, Cherchneff2010}; \citealp{Marassi2014, Marassi2015}; \citealp{Sarangi2013, Sarangi2015}; \citealp{Bocchio2016}; \citealp{Sluder2017}). 
\newline
Once created by stellar sources and injected into the ISM, these grains could evolve by physical processes capable of destroying or growing them.  
The key mechanisms responsible for dust destruction are collisions between grains, SN shocks and thermal sputtering (\citealp{Draine&Salpeter1979}; \citealp{McKee1987}; \citealp{Jones1996}, \citealt{Hirashita2010}; \citealp{Jones2011}; \citealp{Bocchio2014}). 
Grains may also grow by accretion  of gas-phase metals within dense molecular clouds (\citealp{Draine1990}; \citealp{Dominik&Tielens1997}; \citealp{Hirashita2012}; \citealp{Inoue2011}; \citealp{Kohler2015}), thus increasing the dust mass.
\newline
Although the efficiency or even the physical nature of these processes is still highly debated  (\citealp{Ferrara2016}, but see also \citealp{Zhukovska2016}), grain growth has been often invoked by galactic chemical evolution models as the dominant process responsible for production of the dust mass inferred from observations of local (\citealp{Zhukovska2014}; \citealp{deBennassuti2014}; \citealp{Schneider2016}; \citealp{McKinnon2016}; \citealp{Popping2017}; \citealp{Gioannini2017}) and high-redshift galaxies (\citealp{Rowlands2014}; \citealp{Michalowski2015}; \citealp{Mancini2015, Mancini2016}; \citealp{Watson2015}; \citealp{Aoyama2017}; \citealp{Knudsen2017}; \citealp{Wang2017}).
Moreover strong indications for grain growth acting in the ISM come from sub-millimeter (submm) observations of high redshift quasar host galaxies (\citealp{Valiante2011}; \citealp{Mattsson2011}; \citealt{Pipino2011}; \citealp{Calura2014}; \citealp{Valiante2014})  and gamma-ray burst Damped Lyman-alpha Absorbers (\citealp{Wiseman2017}).
Finally, only chemical evolution models accounting for dust growth in the ISM (e.g. \citealp{Asano2013}; \citealp{Zhukovska2014}; \citealp{deBennassuti2014}; \citealp{Popping2017}) can successfully reproduce the observed trend between dust-to-gas mass ratio ($D/G$) and metallicity (\citealp{RemyRuyer2014, RemyRuyer2015}), or can explain the gas surface density dependence of $D/G$ observed in the Magellanic Clouds by IRAS and Planck \citep{RomanDuval2017}.

In this Letter we investigate the efficiency of stellar mechanisms of dust production, aiming at interpreting the dust mass budget of the Milky Way (MW) and the relations between the dust mass ($ M_{\rm dust}$), stellar mass ($M_{\rm \star}$), $D/G$ and metallicity ($Z$) observed for a wide sample of galaxies, spanning $\sim 2$ dex in $Z$ and $\sim 5$ dex in $ M_{ \rm \star}$ (see Sec.~\ref{sec:sample}).
\newline
For this reason, we have implemented different models of dust formation from stellar sources (e.g. \citealp{Ferrarotti&Gail2006}; \citealp{Bianchi&Schneider2007}; \citealp{Zhukovska2008}; \citealp{DellAgli2015, DellAgli2017}; \citealp{Marassi2015}; \citealp{Bocchio2016}) in the latest release of the \texttt{GAMESH} pipeline (\citealp{Graziani2015}), capable of reproducing the stellar, gas and metal mass observed in the MW as well as the 
fundamental scaling relations in the redshift range $\rm 0<z<4$ (\citealp{Graziani2017}). 
\newline
The Letter is organized as follows.
In Sec.~\ref{sec:sample} we briefly describe the sample of data used for the comparisons with our results. 
In Sec.~\ref{sec:model} we summarize the properties of the \texttt{GAMESH} pipeline, and we describe the adopted stellar yields of dust production and their modelling.
Sec.~\ref{sec:results} and Sec.~\ref{sec:discussion} show the results of this work and a discussion of their astrophysical implications.

\section{SAMPLE AND OBSERVATIONS}\label{sec:sample}

Two samples of local galaxies observed with \textit{Herschel} have been used for comparison with our models:  the Dwarf Galaxy Survey (DGS, \citealp{Madden2013})	and the Key Insights on Nearby Galaxies: a Far-Infrared Survey with \textit{Herschel} (KINGFISH, \citealp{Kennicutt2011}).  
These samples are extensively described in \cite{RemyRuyer2014, RemyRuyer2015}. 
The DGS is a sample of 48 star-forming dwarf galaxies with low metallicity ranging from $\rm12 + log(O/H) = 7.14$ to $8.43$.  
Stellar masses span $\sim$4 dex, from $3\times\,10^6 M_{\odot}$  to $\sim 3\times \,10^{10} M_{\odot}$.
The KINGFISH sample contains 61 galaxies with metallicity in the range $\rm12 + log(O/H) = 7.54 - 8.77$ and stellar masses in the range $[2\times10^7 - 1.4\times10^{11}]\, M_{\odot}$, and it probes more metal-rich and massive environments.
The metallicities of the DGS galaxies have been derived using empirical strong emission-line methods (\citealp{Madden2013}), and DGS dust masses have been computed by \cite{RemyRuyer2015}. 
DGS stellar masses were taken from \cite{Madden2014}, who used the prescription of \cite{Eskew2012} and the IRAC 3.6 $\mu$m and 4.5 $\mu$m luminosities.
KINGFISH metallicities have been taken from \cite{Hunt2016}, recalibrated to the \cite{Pettini2004} strong-line $N$2 calibration, and dust masses taken from Hunt et al. (2017) (in preparation) using the photometry presented by \cite{Dale2017}. 
Stellar masses for KINGFISH are taken from \cite{Hunt2016}, computed from the IRAC 3.6 $\mu$m luminosities according to \cite{Wen2013}.
Gas masses for both the DGS and the KINGFISH samples are obtained by combining the contribution of HI (see \citealp{Draine&Li2007} for the KINGFISH galaxies and \citealp{Madden2013} for DGS) and molecular gas, derived through CO measurements (\citealp{RemyRuyer2015}).
\newline
To compare our predictions with the dust mass budget of the MW, observed values from \cite{Planck2011} and \cite{Bovy2013} (where the measured gas mass is converted into dust mass using a standard $(D/G)_{\rm MW} \sim 1/100$; \citealp{Draine&Li2007}) have been adopted.

\section{MODEL DESCRIPTION}\label{sec:model}

The cosmological simulation has been performed with the latest release of the \texttt{GAMESH} pipeline \citep{Graziani2015, Graziani2017} which combines a high resolution N-Body simulation\footnote{The N-Body simulation is based on the code $\rm GCD+$ (\citealp{Kawata2013}) and adopts a flat $\rm \Lambda CDM$ cosmology with $\rm \Omega_m = 0.32$, $\rm \Omega_\Lambda = 0.68$, $\rm \Omega_b = 0.049$ and $h =0.67$ (\citealp{Planck2014}).}, a novel version of the semi-analytic, data-constrained model \texttt{GAMETE} (\citealp{Salvadori2007}) and the third release of the radiative transfer code \texttt{CRASH} (\citealp{Graziani2013}) computing gas ionization through hydrogen, helium and metals.

\texttt{GAMESH} follows the cosmological evolution of a cubic volume of 4~cMpc side length, centered on a well resolved MW-like halo\footnote{The MW-like halo has a mass $\rm M_{MW}=1.7\times10^{12}M_\odot$ at $z=0$ and it is resolved with a dark matter particle resolution mass of $\rm 3.4\times 10^5 M_\odot$.}, by accounting for star formation, chemical enrichment, Pop III/Pop II transition and SN-driven feedback. 
Comparison with recent observations of candidate MW progenitors at $0<z<2.5$ reproduce the galaxy main sequence, the mass-metallicity relation and the fundamental plane of metallicity relations at $0<z<4$ (\citealp{Graziani2017}). The interested reader can find more details in \cite{Graziani2015, Graziani2017} and \cite{Schneider2017}. 

The present work adopts a novel extension of \texttt{GAMESH} in which dust formation by stellar sources is computed, moving its semi-analytic scheme towards more advanced release of \texttt{GAMETE} \citep{Valiante2014,deBennassuti2017}. For the purpose of this study, we neglect dust destruction by interstellar shocks and the only physical mechanism that can decrease the dust content of the
ISM is astration. To maintain the flexibility of the \texttt{GAMESH} pipeline, the current release runs on different theoretical models of stellar dust yields, that we 
briefly describe below.   

\subsection{Modelling dust production yields}\label{sec:dustmodel}

AGB and SNe have been considered as the two major dust producers. Their relative 
importance depends on the stellar initial mass function (IMF), on the star formation history, and on the
mass and metallicity dependence of the stellar dust yields \citep{Valiante2009}. 
Using AGB dust yields from \citet{Ferrarotti&Gail2006} and \citet{Zhukovska2008} (hereafter, Z08)
and SN dust yields from \citet{Bianchi&Schneider2007} (hereafter BS07), \citet{Valiante2009} showed that 
AGB stars can contribute and eventually dominate dust enrichment on relatively short evolutionary 
timescales (150 - 200 Myr) when the stars are assumed to form in a burst with a Salpeter-like IMF.
Recently, a new grid of AGB dust yields for stars with masses in the range $[1-8] M_\odot$ and metallicity $0.01 Z_\odot \leq Z \leq Z_\odot$
has been computed by \citet{ventura_etal_2012a, ventura_etal_2012b,ventura_etal_2014}, \citet{dicriscienzo_etal_2013}, \citet{DellAgli2017}.
These are based on numerically integrated stellar models by means of the ATON code and predict a different 
mass and metallicity dependence of dust production rates with respect to Z08\footnote{We refer to the original papers for a thorough discussion
of the models and a comparison with Z08 models.}. As a result, assuming the stars to form in a single burst with a Salpeter-like IMF and 
adopting these new AGB dust yields (hereafter ATON yields), \citet{Schneider2015} 
has found that when the initial metallicity of the stars is $Z \leq 0.2 Z_\odot$, the contribution
of AGB stars is always sub-dominant with respect to that of SNe and dominate on a timescale $\sim$ 500~Myr only at higher metallicity.

The above conclusion depends on the adopted SN yields and on the fraction of freshly formed dust that gets
destroyed by the passage of the SN reverse shock generated by the interaction of the SN blast wave with its surrounding medium 
and propagating through the ejecta. 
This shock triggers the destruction of SN-condensed dust grains 
through thermal sputtering due to the interaction of dust grains with particles in the gas,
sublimation due to collisional heating to high temperatures, and vaporisation of part of the colliding grains during grain-grain collisions
\citep{Nozawa2007,Bianchi&Schneider2007, Silvia2010, Silvia2012, Marassi2014, Marassi2015, Bocchio2016, Micelotta2016}.
Here we explore different combinations of SN dust yields and reverse shock destruction efficiencies. 
We consider the metallicity- and mass-dependent SN dust yields computed by BS07 applying standard nucleation
theory to the grid of SN explosion models by \citet{Woosley&Weaver1995} with progenitor masses in the range $[12-40]~M_{\odot}$, metallicity
$[10^{-4} - 1]~Z_\odot$ and adopting a constant explosion energy of $1.2\times 10^{51}$ ergs. When no reverse shock is considered, the predicted
SN dust yields are in the range $[0.1 -0.6]~M_\odot$. We also consider the effects of reverse shock destruction, adopting the model where the
circum-stellar medium density is $\sim 1$~cm$^{-3}$, which leads to $\approx 10\%$ smaller dust yields. When compared to \textit{Herschel} data of
young SN remnants, these two sets of yields appear to brackets the inferred dust masses \citep{Schneider2015}. 

For this study we also adopt new SN dust yields spanning a larger range of 
progenitor masses and metallicity (Marassi et al.~2017, hereafter MS17).
In particular we consider the grid of {\itshape calibrated} SN models, 
where the explosion energy is not fixed a priori (\citealp{Limongi2017}) 
but is instead calibrated requiring the ejection of a specific amount of radioactive $\rm ^{56}Ni$ for each SN progenitor.
The amount of $\rm ^{56}Ni$ for each SN progenitor is obtained from the best fit 
of the observations (Marassi et al. 2017, in prep.). For these SN dust yields,
we assume that between [1 - 8]\% of the original dust mass is able to survive the
passage of the reverse shock, contributing to dust enrichment. These values have
been calibrated on the results obtained by \citet{Bocchio2016}, where the new 
code GRASH-Rev\footnote{GRASH-Rev follows the dynamics of dust grains 
in the shocked SN ejecta and computes the time evolution of the mass, composition
and size distribution of the grains \citep{Bocchio2016}.} has been applied to four
SN models selected to best-fit the properties of SN 1987A,
CasA, the Crab nebula and N49, that have been observed with both \textit{Spitzer} and \textit{Herschel}. 
This study (e.g. \citealp{Bocchio2016}) suggests that the largest dust mass destruction is predicted to occur between $10^3$ and $10^5$ yr
after the explosion. As a result, since the oldest SN in the sample has an age of 4800 yr, the observed dust
mass can only provide an upper limit to the mass of SN dust that will enrich the ISM.

\section{RESULTS}\label{sec:results}

Here we show the results of simulations where we adopt different combinations of dust production models by stellar sources.
When considering SN contributions, we discuss separately the two cases with (RS) and without (nRS) the effects of the reverse shock.
In Sec.~\ref{sec:MW} we compare the observed dust mass of the MW with masses arising from different models of dust production 
in our simulations. In Sec.~\ref{sec:galaxies} we extend the comparison to a distribution of local galaxies spanning a wide range in 
metallicity and stellar masses.

\subsection{MW dust mass assembly}\label{sec:MW}

\begin{figure}
	\centering
	\includegraphics[width=1\columnwidth]{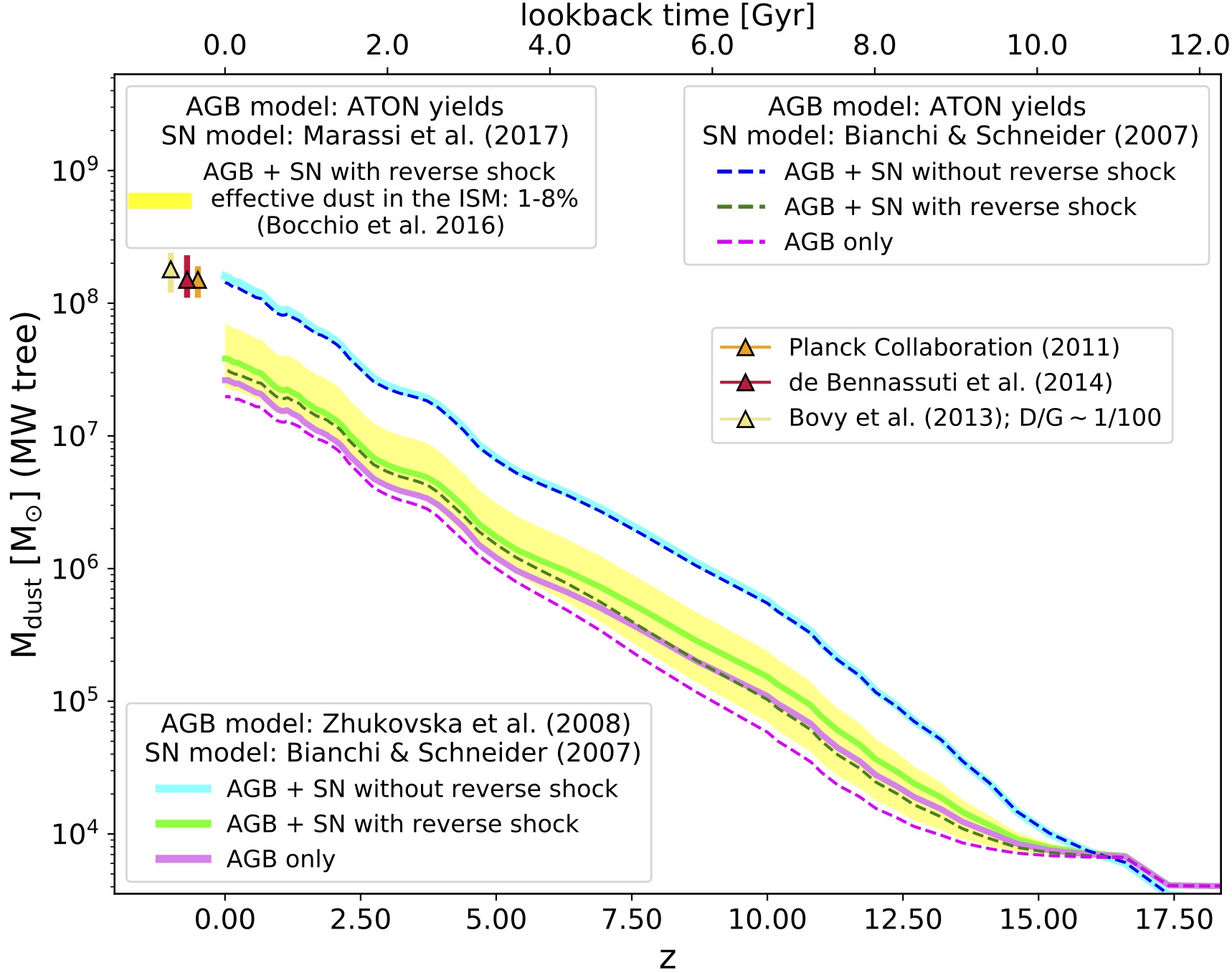}
	\caption{The dust mass evolution of the MW-sized halo as a function of redshift and lookback time. 
	    Different line styles correspond to different models of dust production by stellar sources: 
		solid lines (dashed lines) refer to SN dust yields by BS07 and  AGB dust yields by Z08 (ATON).
		Different line colours discriminate RS and nRS models (see the legenda). An AGB-only model is also explored (pink lines).
		The shaded yellow region correspond to the new M17 (SN) + ATON (AGB) model.  
		The triangular points represent the MW dust mass at $z=0$, as inferred by observations from \protect\cite{Planck2011} (orange) and \protect\cite{Bovy2013} (yellow, in this case the measured gas  mass has been converted in dust mass assuming a standard $D/G\rm _{MW}\sim1/100$) and simulations (\citealp{deBennassuti2014}).}
	\label{fig:dustMW}
\end{figure}

The dust mass assembly in the MW-sized halo is shown in Fig.~\ref{fig:dustMW} as a function of redshift and lookback time. 
Violet (solid/dashed) lines refer to the amount of dust produced by AGB stars only: it is immediately evident that, independently of the adopted yields, 
these stars alone cannot produce more than $\sim3\times 10^7 M_{\odot}$ of dust, with a discrepancy of $\sim30\%$ between old (Z08) and new (ATON) models. 
The higher dependence on metallicity of ATON yields leads to lower dust masses at $z\gtrsim5$.
At all redshifts, SNe dominate dust production, even in the RS case. The comparison of model predictions with observations 
shows that the existing dust mass ($M_{\rm dust} \sim 1.5\times 10^8 M_{\odot}$) in the MW can be produced by stellar dust sources only if
all the dust formed in AGB and SN ejecta is injected in the ISM without suffering destruction by the RS (cyan
and blue dashed lines). We consider this to be an unrealistic assumption. In fact, \citet{Bocchio2016} show that while SN 1987A is too young for the reverse
shock to have affected the dust mass, in Cas A, Crab and N49 the reverse shock has already destroyed between 10 - 40\% of the initial dust mass, 
despite the relatively young age of these SN remnants.
\newline
On the other hand, when the reverse shock effect is included, the dust mass produced by stellar sources and effectively injected into the ISM (i.e. the grains which survive the passage of the shock) is a factor $\sim$4  lower than the observed value, as shown by Fig.~\ref{fig:dustMW}. Here the yellow area shows the M17 with RS models (with the upper and lower
boundaries corresponding to 1 - 8\% destruction), in good agreement with predictions from the old stellar yields of BS07 with RS.

We conclude that, independently of the dust production yields, a model where dust has only a stellar origin fails at reproducing the total dust mass in the MW. 
As other metallicity-related relations are correctly predicted by the simulation \citep{Graziani2017}, we interpret this result as a clear 
indication that dust evolution in the MW ISM plays an important role. 
Moreover, as both dust destruction (likely acting in the hot ISM phase) and grain growth in the cold ISM are missing in the present model, 
Fig.~\ref{fig:dustMW} suggests an efficient  process of dust growth, capable of compensating both the missing factor $\sim4$ commented above, 
and the additional effects of dust sputtering and destruction acting in the hot medium of the MW halo.  

\subsection{Dust abundances in galaxies at $z\sim 0$}\label{sec:galaxies}

\begin{figure}
	\centering
	\includegraphics[width=1\columnwidth]{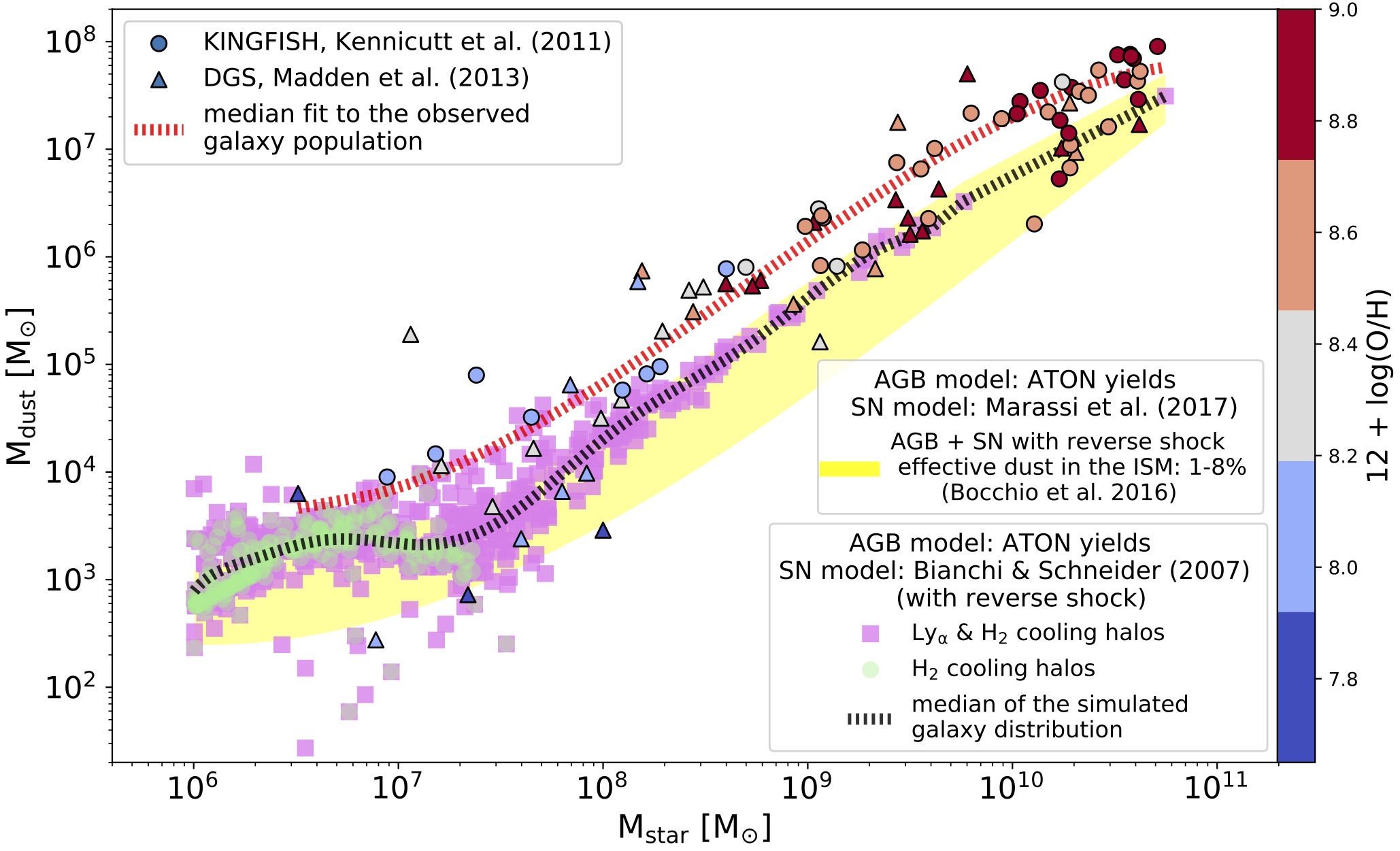}
	\caption{The dust mass of simulated galaxies (pink squares) at $z = 0$ as a function of their stellar mass. 
	Green dots represent the sub-sample of galaxies with virial temperature $\rm T_{vir}<2\times10^{4}K$. 
	The black dashed line indicates the median trend of the simulated galaxy distribution.
	Here, ATON + BS07 (with RS)  and ATON + M17 (with RS) models for dust production by stellar sources are adopted.
	Observational points of galaxies from the DGS (triangles) and KINGFISH (circles) surveys are shown, colour coded for different values of their metallicity. 
	The red dashed line indicates the median trend of the observed galaxy distribution.
	}
	\label{fig:dust_VS_star}
\end{figure}

Here we extend our analysis by comparing in Fig.~\ref{fig:dust_VS_star} the $M_{\rm dust} - M_{\rm \star}$ relation of simulated galaxies at $z = 0$ with local observations from the DGS and KINGFISH surveys (see Sec.~\ref{sec:sample}). 
Hereafter we adopt the ATON yields for AGB stars and SN yields by the BS07 and by MR17 with RS.
The symbols indicating the observed galaxies are colour coded (see color palette in the figure) for different values of their metallicity. 
Fig.~\ref{fig:dust_VS_star} shows that the dust masses predicted by the simulation are (on average) systematically lower (a factor $\sim3$) than the observed ones, at any given stellar mass.
Such a discrepancy decreases towards the low-$ M_{\rm \star}$ tail of the distribution, where observed galaxies are mostly metal poor ($\rm 12 + log(O/H)\lesssim8$). 
This result confirms - on a statistical sample of local galaxies - that dust produced by stellar sources only is insufficient to account for the observed mass.
Depending on the efficiency of the reverse shock model, the new SN models seem to largely under-predict the observed trends (see yellow area), shifting the predicted median at lower dust masses, at fixed $M_{\rm \star}$. 

\begin{figure}
	\centering
	\includegraphics[width=1\columnwidth]{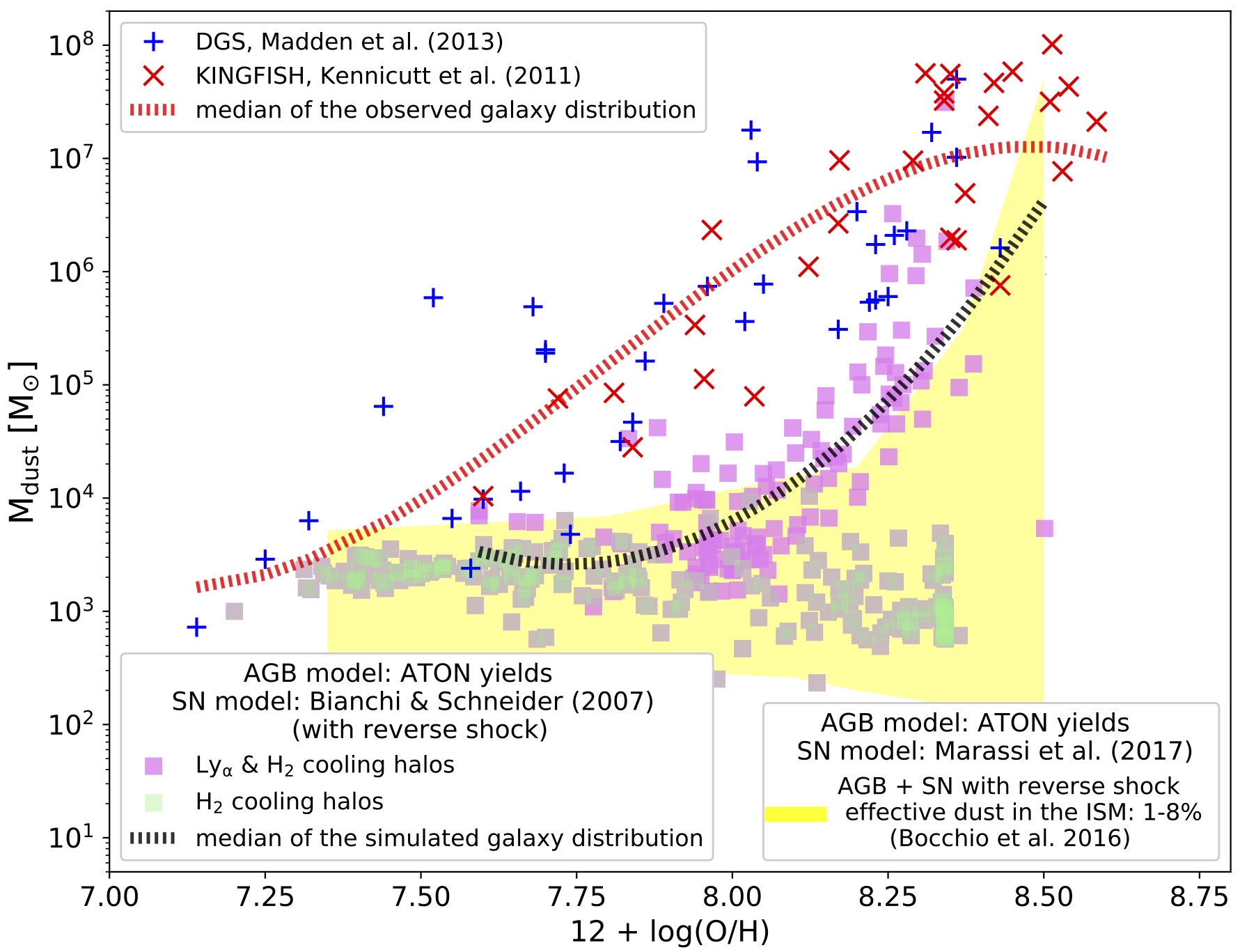}
	\caption{Dust mass as a function of metallicity. Lines and symbols are the same as in Fig.~\ref{fig:dust_VS_star} but observations
	 from the DGS and KINGFISH surveys are shown as blue plus and red crosses. 
	The black dashed line indicates the median trend of the observed galaxy distribution.	
	}
	\label{fig:dust_VS_Z}
\end{figure}

Similar considerations arise from Fig.~\ref{fig:dust_VS_Z} where we show the  $M_{\rm dust} - Z$ relation. 
The observed dust mass-metallicity relation is largely under-predicted by the simulations, both for galaxies hosted in Ly$\alpha$-cooling and in H$_2$-cooling halos. 
In addition, the comparison of the two dashed lines show that the simulated galaxies fail to reproduce the observed average trend.

\begin{figure}
	\centering
	\includegraphics[width=1\columnwidth]{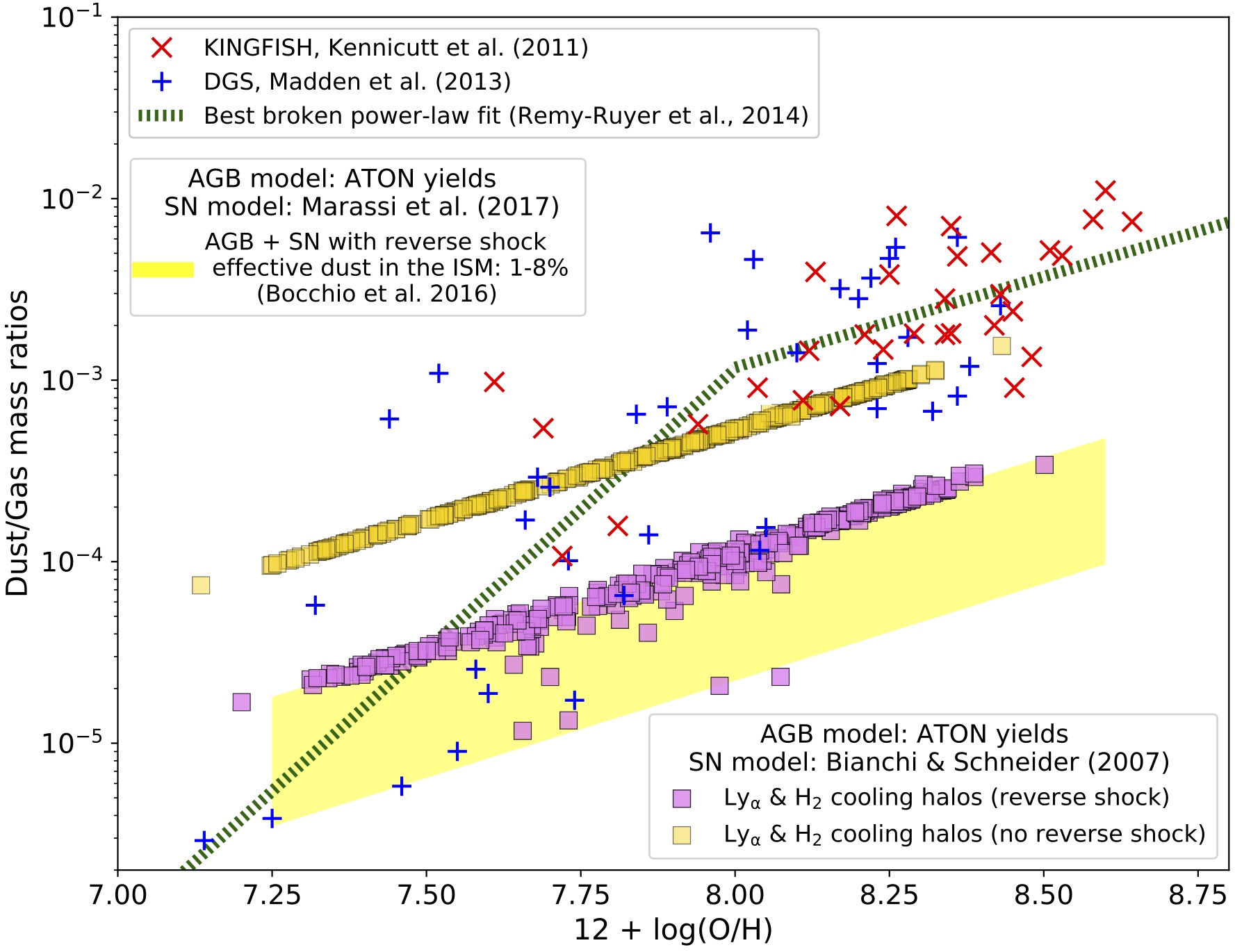}
	\caption{The $D/G$ of simulated galaxies at $z \sim 0$  as a function of their metallicity. Lines and symbols are the same as in Fig.~\ref{fig:dust_VS_Z}.
	For comparison, we also show the ATON + BS07 model with nRS (yellow squares) and the broken-power law fit to the observed galaxy distribution computed 
	in \protect\cite{RemyRuyer2014} (green dashed line).
	}
	\label{fig:dtg_VS_Z}
\end{figure}

Finally Fig.~\ref{fig:dtg_VS_Z} shows the $D/G$ of simulated galaxies at $z \sim 0$  as a function of their metallicity. 
Observational data from the DGS and KINGFISH surveys are reported, and the broken-power law fit to the 
$D/G-Z$ relation, computed by \cite{RemyRuyer2014}, is also shown as the dashed line. In addition to the
reference models considered in the previous figures, here we also show the ATON + BS07 model with nRS (yellow squares). 
In models where the SN dust yields are reduced by the effect of the RS, the $D/G$ of the simulated galaxies 
confirm our previous conclusions, being about one order of magnitude lower than the observed one at intermediate-high metallicity.
Such a discrepancy tends to decrease in the low-$Z$ tail of the galaxy distribution ($\rm 12 + log(O/H)\lesssim8$), where the observed $D/G - Z$ enters a new regime (i.e. the slope changes).
When nRS is considered, the predicted $D/G$ of simulated galaxies at high-metallicity are closer to the observed values. However, regardless of the 
adopted stellar yields, none of the models is able to reproduced the observed double power law trend of the  $D/G-Z$ relation (\citealp{Hirashita2012}; \citealp{Asano2013}; \citealp{Zhukovska2014}).

	
\section{Discussion and Conclusions}\label{sec:discussion}
	
Our results confirm that dust evolution models accounting for \textquoteleft stellar production-only\textquoteright\, fail at reproducing a number of observables. 
In particular:
\begin{enumerate}
	\item the dust mass produced by stars in the MW and effectively injected into the ISM after surviving the passage of the SN reverse shock is predicted to be a factor $\sim4$ lower than observed, with the uncertainty depending on which model of dust formation by AGB stars is adopted (see Fig.~\ref{fig:dustMW}).
	
	\item the predicted dust mass budget of the simulated population of local galaxies results to be, on average, systematically underestimated with respect to observations, with visible consequences arising when comparing our simulated $ M_{\rm dust} - M_{\rm \star}$  (Fig.~\ref{fig:dust_VS_star}), $ M_{\rm dust}-$$Z$ (Fig.~\ref{fig:dust_VS_Z}) and $D/G-Z$ (Fig.~\ref{fig:dtg_VS_Z}) relations with the observed trends.
	Nevertheless, there are insights for such a discrepancy declining in proximity of the low-$Z$ and low-$M_{\rm \star}$ tails of the galaxy distribution.
\end{enumerate}

Hence the amount of dust injected into the ISM by SN explosions (accounting for the destructive effect of the reverse shock) and AGB stellar winds is not sufficient to reproduce the observed mass of interstellar dust.
The inclusion of dust destruction by interstellar shocks, that we have neglected, would further strengthen this conclusion.
\newline
For completeness, we tested the hypothesis that dust grains formed in SN ejecta are injected in the ISM without being affected by the passage of the reverse shock.
While this unrealistic model succeeds at reproducing the observed dust mass of the MW (Fig.~\ref{fig:dustMW}), it fails at reproducing the observed $D/G-Z$ relation 
in the low-$Z$ tail of the local galaxy population.

Altogether our results suggest that additional (non-stellar) mechanisms of dust growth are at play during the galaxy evolution across cosmic times.
Our findings are consistent with a number of previous works  (see Sec.~\ref{sec:introduction}) where a significant grain growth in the dense phase of the ISM has been invoked as a supplementary process to explain the rapid dust enrichment observed in high-z objects (\citealp{Valiante2011, Valiante2014}; \citealp{Mattsson2011}; \citealt{Pipino2011}; \citealp{Calura2014}; \citealp{Michalowski2015}; \citealp{Mancini2015, Mancini2016}), as well as the observed 
$D/G-Z$ trend in local galaxies (\citealp{RemyRuyer2014}; \citealp{Zhukovska2014}; \citealp{deBennassuti2014}; \citealp{Schneider2016}; \citealp{Popping2017}) 
and the strong gas density dependence of the $D/G$ observed in the Magellanic Clouds (\citealp{RomanDuval2017}). 
Our work strengthens these previous findings by showing that these conclusions are largely independent of the adopted dust yields and reverse shock modelling.
\newline
A natural consequence of these results is that data-calibrated models of the ISM in galaxy evolution must consider additional channels responsible for both dust formation in the dense phase and dust destruction in the hot phase. 
A deeper understanding of these phenomena is deferred to future works, where an improved version of the \texttt{GAMESH} pipeline, providing for an accurate modelling of the multi-phase ISM, will be exploited.

\section*{Acknowledgements}

The research leading to these results has received funding from the European Research Council under the European Union’s Seventh Framework Programme (FP/2007- 2013)/ERC Grant Agreement n. 306476. FD acknowledges support provided by the MINECO grant AYA−2014−58082-P.




\bibliographystyle{mnras}
\bibliography{biblio} 


%
%


\bsp	
\label{lastpage}
\end{document}